\documentclass{article}

    \PassOptionsToPackage{numbers, compress}{natbib}

\usepackage[preprint]{neurips_2024}

\usepackage[utf8]{inputenc} 
\usepackage[T1]{fontenc}    
\usepackage{hyperref}       
\usepackage{url}            
\usepackage{booktabs}       
\usepackage{amsfonts}       
\usepackage{nicefrac}       
\usepackage{microtype}      
\usepackage{epigraph}
\usepackage{array}

\usepackage{graphicx}
\graphicspath{ {../figs/} }
\usepackage{caption}        
\usepackage{sidecap}        
\usepackage{subcaption}

\usepackage[normalem]{ulem}	
\usepackage[toc,page]{appendix}	
\usepackage{blindtext}
\usepackage{setspace}
\usepackage{amsmath}
\usepackage{mathtools}
\usepackage{booktabs}
\usepackage{hyperref}
\usepackage{enumitem}
\usepackage{nicefrac}       

\makeatletter
\DeclareRobustCommand\onedot{\futurelet\@let@token\@onedot}
\def\@onedot{\ifx\@let@token.\else.\null\fi\xspace}

\usepackage{wrapfig}
\usepackage[bottom]{footmisc}
\usepackage{color,soul}
\usepackage{nicematrix,tikz} 
\usepackage{ulem}  
\setcitestyle{square}
\setcitestyle{citesep={,}}
\usepackage{lipsum}
\usepackage{caption}
\usepackage{adjustbox}

\usepackage{amsmath}
\usepackage{amssymb}
\usepackage{bm}
\usepackage{mathtools}
\usepackage{amsthm}

\usepackage{float}
\usepackage{algorithm}
\usepackage{algorithmic}
\usepackage{multirow}

\usepackage[capitalize,noabbrev]{cleveref}

\usepackage[textsize=tiny]{todonotes}

\usepackage{titletoc}
\newcommand\DoToC{%
  \startcontents
  \printcontents{}{1}{\textbf{Appendix Contents}\vskip3pt\hrule\vskip5pt}
  \vskip3pt\hrule\vskip5pt
}

\usepackage{listings}
\lstnewenvironment{rsverbatim}[1][]{%
  \lstset{
    basicstyle=\small\ttfamily,
    columns=flexible,
    breaklines=true,
    frame=tb,
    #1
  }%
}{}

\usepackage{tikz}
\usetikzlibrary{matrix,chains,positioning,arrows, calc}
\usepackage{float}
\usepackage{amsthm}
\usepackage{multirow}
\usepackage{multicol}

\setboolean{use_header_icon}{true}
\ifthenelse{\boolean{use_header_icon}}{%
  \papertitlewithimage{figs/nifty-icon}{NIFTY Financial News Headlines Dataset}
}{%
  \title{NIFTY Financial News Headlines Dataset}
}

\author{%
  Raeid Saqur\\ 
  Department of Computer Science\\
  University of Toronto\\
  \texttt{raeidsaqur@cs.toronto.edu} \\
  \And
  Ken Kato \\
  Department of Computer Science\\
  University of Toronto\\
  \texttt{ken.kato@mail.utoronto.ca} \\
  \AND
  Nicholas Vinden \\
  Department of Computer Science\\
  University of Guelph\\
  \texttt{nvinden@uoguelph.ca} \\
  \And
  Frank Rudzicz \\
  Faculty of Computer Science\\
  Dalhousie University \\
  \texttt{frank@dal.ca} \\
}

\definecolor{Gray}{RGB}{217,234,211}

\newcommand{\niftyurl}{https://huggingface.co/datasets/raeidsaqur/NIFTY}

\begin{document}
\maketitle


\begin{abstract}
We introduce and make publicly available the NIFTY Financial News Headlines dataset, designed to facilitate and advance research in financial market forecasting using large  language models (LLMs). This dataset comprises two distinct versions tailored for different modeling approaches: (i) NIFTY-LM ($\mathcal{D}_{LM}$), which targets supervised fine-tuning (SFT) of LLMs with an auto-regressive, causal language-modeling objective, and (ii) NIFTY-RL ($\mathcal{D}_{RL}$), formatted specifically for alignment methods (like reinforcement learning from human feedback (RLHF)) to align LLMs via rejection sampling and reward modeling. Each dataset version provides curated, high-quality data incorporating comprehensive metadata, market indices, and deduplicated financial news headlines systematically filtered and ranked to suit modern LLM frameworks. We also include experiments demonstrating some applications of the dataset in tasks like stock price movement and the role of LLM embeddings in information acquisition/richness. The NIFTY dataset along with utilities (like truncating prompt's context length systematically) are available on Hugging Face at \href{\niftyurl}{NIFTY Dataset}.
\end{abstract}


\section{Introduction}\label{sec:intro}
Recent advances in deep learning research have significantly changed our approach to solving complex problems in many fields.
GraphCast's~\cite{graphcast_lam2022graphcast} success in weather forecasting, and AlphaFold's~\cite{alphafold} breakthrough success in 3D protein structure prediction are two emblematic examples of modern machine learning (ML) successes that have supplanted or radically shifted decades-old (complex, heuristic) approaches. 
Financial market forecasting is a hard problem, arguably more complex than any of the aforementioned issues. While the allure of solving this problem extends well beyond the academic and scientific communities (for obvious, intrinsic rewards), the core problem can be formalized from an RL (or optimal control) perspective by setting up the problem as a (highly complex) partially observable MDP (POMDP)~\cite{kaelbling1998planning}. The true, plausibly large number of variables and mechanics that move the market are hidden or unobservable. Thus, reliable market simulation, thereby generating randomized market value trajectories to train agents in simulation, is not yet effective, making market prediction in essence a \textit{one-shot} learning task with only one true trajectory or available environment history. Any mapping of input observations ($o_t \in \mathcal{O}$) to output price movement (i.e., market/environment reaction) learned via traditional ML techniques does not generalize well to out-of-domain (or regime-shifted) distributions due to the hidden, underlying correlation and covariate shifts in a dynamic market regime~\cite{regime_ang2012regime, regime_guidolin2008size}. Basically, even if we are able to train a model that fits perfectly to past market trajectories (i.e., success in backtesting), it does not guarantee future accuracy.

News headlines are reasonable, albeit extremely abstract, proxies for approximating the underlying factors that move markets. In this work, we contribute a news headlines dataset curated over the past decade in a manner that can be easily consumed by modern LLM models, frameworks, and pipelines; thereby allowing fast and broader research participation, including from the public community.

\paragraph{Contributions} Our main contributions with this work are:
1. The \textbf{NIFTY} \textbf{dataset}: We open-source the large language modelling and preference tuning dataset used for our work. NIFTY, or the \textit{News-Informed Financial Trend Yield}, has the longest coverage of news headlines from the past decade (2010 to 2023). The curated dataset includes topic tags, relevant filtered and deduplicated financial market news, and provides the most comprehensive, high-quality financial research dataset to date for the community.

2. \textbf{Role of LM Embeddings} in the quality of observations or news data: We perform a comprehensive analysis and discussion on the quality of source embeddings in terms of information gain from the corresponding input observations or news data. We present two interesting findings from the analysis: i. the underlying LLM architectural blocks (\textit{encoder}, \textit{decoder}, and \textit{encoder-decoder}) of models matter, and ii. model size matters, i.e., the rich get richer since embeddings from larger LMs (by parameter size) tend to have better \textit{information gain} compared to smaller-variants and architectures of the same family~\S\ref{app:embeddings}.

\paragraph{Organization} The rest of the paper is organized as follows: \S\ref{sec:dataset} details the NIFTY dataset, \S\ref{sec:applications} discusses and proposes a few uses and applications in research using the contributed dataset, and \S\ref{sec:experiments} presents selected experiments and baseline results for aiding future research.

\section{NIFTY Financial News Headlines Dataset}\label{sec:dataset}
The News-Informed Financial Trend Yield (NIFTY) dataset is a processed and curated daily news headlines dataset for the stock (US Equities) market price movement prediction task. NIFTY is comprised of two related datasets, NIFTY-LM and NIFTY-RL. In this section we outline the composition of the two datasets, and additional details.

\paragraph{ Dataset statistics }
Table~\ref{table:NIFTY-stats} and Table~\ref{table:NIFTY-date-ranges} present pertinent statistics related to the dataset.

\setlength{\tabcolsep}{2pt} 
\renewcommand{\arraystretch}{1.2}
\begin{table}[ht]
\centering
\begin{minipage}[t]{0.48\textwidth}
    \centering
    \caption{Statistics and breakdown of splits sizes}
    \label{table:NIFTY-stats}
    \vspace{0.5em}
    \begin{adjustbox}{width=\textwidth}
    \begin{tabular}{lc}
    \toprule
    Category & Statistics \\
    \midrule
    Number of data points & 2111 \\
    Number of Rise/Fall/Neutral label & 558 / 433 / 1122 \\
    Train/Test/Evaluation split & 1477 / 317 / 317 \\
    \bottomrule
    \end{tabular}
    \end{adjustbox}
\end{minipage}%
\hfill
\begin{minipage}[t]{0.48\textwidth}
    \centering
    \caption{Date Ranges of news headlines in splits}
    \label{table:NIFTY-date-ranges}
    \vspace{0.5em}
    \begin{adjustbox}{width=\textwidth}
    \begin{tabular}{lcc}
    \toprule
    Split & Num. Samples & Date range \\
    \midrule
    Train & 1477 & 2010-01-06 to 2017-06-27 \\
    Valid & 317 & 2017-06-28 to 2019-02-12 \\
    Test & 317 & 2019-02-13 to 2020-09-21 \\
    \bottomrule
    \end{tabular}
    \end{adjustbox}
\end{minipage}
\end{table}

\subsection{Methodology: Data Procurement and Aggregation}
The dataset aggregates high-quality news headlines that are related to the financial market movement. For example, the headline: ``Justin Trudeau gets divorced'', while dominant, is not germane or likely to influence the stock market. We first aggregated news headlines from various news sources by internet scraping from accessible sites. Then we fed the aggregated news through a textual topic modeling model trained to rank and isolate headlines related to financial topics. Our pipeline then deduplicates, filters and ranks the headlines per day. Finally, we select the top headlines from the ranked list that respects the prompting `context length' of LLMs (we tried to accommodate Llama 2 class models and above), although the context length is no longer a limiting factor for SoTA LLMs. 

\begin{figure}[!t]
\begin{center}
\centerline{\includegraphics[width=\textwidth]{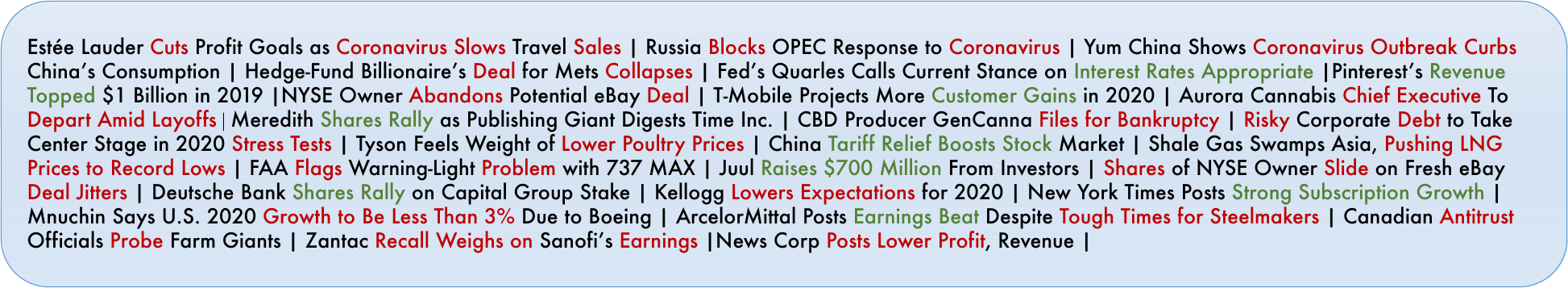}}
\caption{A snapshot of the `news` key value on date: 2020-02-06, at the upstart of the global coronavirus epidemic. Our $\pi_{LM}$ policy's prompt is composed of task instruction as query prefix, market context, and this news value concatenated: $s.t.$ 
$x_p \leftarrow (x_{instruction}; x_{context}; x_{news})$. The semantic text colors \textcolor{red}{red}, and \textcolor{green}{green} conveys negative and positive sentiments. The day's market relevant news was dominated by mostly negative sentiments. }
\label{fig:nifty-news}
\end{center}
\vskip -0.2in
\end{figure}

\subsection{Dataset Structure}\label{ssec:nifty_dataset}
Each dataset split corresponds to a `\texttt{jsonl}` file containing each day's record as a JSON object line. Each JSON-object line of the $\mathcal{D}_{LM}$ contain high-quality, processed (one-turn) conversational query, where a query $x_q$ comprises of a prompt $x_p$ and a response $x_r$, i.e.,  \( x_q = (x_p; x_r) \). Thus, this dataset samples can be used for supervised fine-tuning (SFT) of a pretrained LM policy using the language modeling objective. 
Similarly, the \texttt{NIFTY-RL} dataset compiles a preferences dataset for rejection sampling and RL fine-tuning availing samples of chosen and rejected labels, as shown in Equation~\eqref{eq:rl-dataset}:

\begin{equation}
\mathcal{D}_{RL} = \left\{ \left(x^{(i)}_p, x_{r_w}^{(i)}, x_{r_l}^{(i)}\right) \right\}_{i=1}^{N} \quad \text{where} \quad (x_{r_w} \succ x_{r_l} \mid x_p)
\label{eq:rl-dataset}
\end{equation}

\begin{figure}[ht]
\centering
    \begin{subfigure}{\columnwidth}
      \centering
      \includegraphics[width=\linewidth]{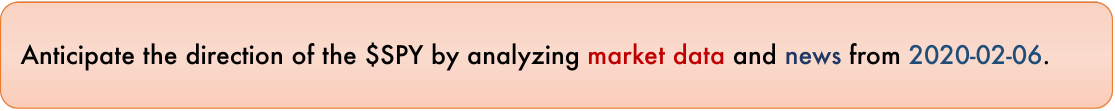}
      \caption{Instruction component of a $\pi_{LM}$ policy query $x_q$.}
      \label{fig:nifty-question}
    \end{subfigure}
\vskip 0.1in
    \begin{subfigure}{\columnwidth}
      \centering
      \includegraphics[width=\linewidth]{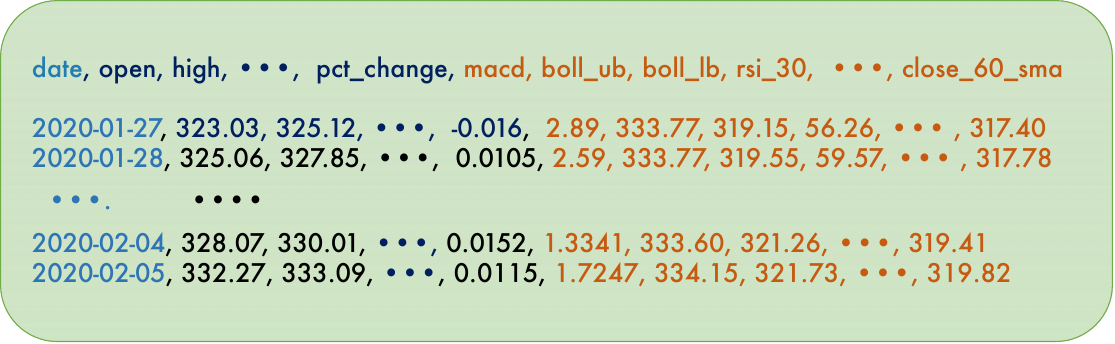}
      \caption{The market's \textbf{history} is provided as the past $t$ days of numerical statistics like the (OHLCV) price (in blue) and common technical indicators (in orange) (e.g. moving averages) data.}
      \label{fig:nifty-context}
    \end{subfigure}
\caption{Breaking down the instruction or prompt prefix, and market context components of a prompt, $x_p$.}
\label{fig:nifty-query}
\vskip -0.2in
\end{figure}

\subsection{NIFTY-LM: SFT Fine-tuning Dataset}


The NIFTY-LM prompt dataset was created to finetune and evaluate LLMs on predicting future stock movement given previous market data and news headlines. The dataset was assembled by aggregating information from various internet news sources from January 6, 2010, to September 21, 2020 -- including headlines from The Wall Street Journal and Reuters News, as well as market data of the \$SPY index from Yahoo Finance. The NIFTY-LM dataset consists of:

\setitemize{label=\textbullet, leftmargin=*}
\begin{itemize}
    \item \textbf{Meta data}: Dates and data ID.
    \item \textbf{Prompt ($x_p$)}: LLM question ($x_{question}$), market data from previous days ($x_{context}$), and news headlines ($x_{news}$).
    \item \textbf{Response}: Qualitative movement label ($x_r$) $\in{} \{Rise, Fall, Neutral\}$, and percentage change of the closing price of the \$SPY index.
\end{itemize}

To generate LLM questions, ($\boldsymbol{x_{question}}$),  we followed the self-instruct \cite{wang2023selfinstruct} framework where we used the OpenAI GPT-4 model to create 20 variations of the instruction below:

\begin{quote}
Create 20 variations of the instruction below. \newline
Examine the given market information and news headlines data on DATE to forecast whether the \$SPY index will rise, fall, or remain unchanged. If you think the movement will be less than 0.5\%, then return 'Neutral'. Respond with Rise, Fall, or Neutral and your reasoning in a new paragraph.
\end{quote}

Where \texttt{DATE} would be substituted later, during the training phase with a corresponding date.

\paragraph{Context} The key `context' ($\boldsymbol{x_{context}}$) was constructed to have newline delimited market metrics over the past T ($\approx$ 10) days (N.B. Not all market data for the past days for were available and therefore prompts might have less than 10 days of market metrics.). 

Table~\ref{tab:dataset_columns} show the details of financial context provided in each day's sample.

\begin{table}[h]
\centering
\caption{Summary of the dataset columns with their respective descriptions.}
\label{tab:dataset_columns}
\begin{adjustbox}{width=\columnwidth,center}
\begin{tabular}{@{}ll@{}}
\toprule
\textbf{Column Name}                  & \textbf{Description}                                                     \\ \midrule
Date                                  & Date of the trading session                                              \\
Opening Price                         & Stock's opening market price                                             \\
Daily High                            & Highest trading price of the day                                         \\
Daily Low                             & Lowest trading price of the day                                          \\
Closing Price                         & Stock's closing market price                                             \\
Adjusted Closing Price                & Closing price adjusted for splits and dividends                         \\
Volume                                & Total shares traded during the day                                       \\
Percentage Change                     & Day-over-day percentage change in closing price                         \\
MACD                                  & Momentum indicator showing the relationship between two moving averages \\
Bollinger Upper Band                  & Upper boundary of the Bollinger Bands, set at two standard deviations above the average \\
Bollinger Lower Band                  & Lower boundary, set at two standard deviations below the average         \\
30-Day RSI                            & Momentum oscillator measuring speed and change of price movements       \\
30-Day CCI                            & Indicator identifying cyclical trends over 30 days                      \\
30-Day DX                             & Indicates the strength of price trends over 30 days                     \\
30-Day SMA                            & Average closing price over the past 30 days                             \\
60-Day SMA                            & Average closing price over the past 60 days                             \\ \bottomrule
\end{tabular}
\end{adjustbox}
\end{table}

\paragraph{News Headlines} $\boldsymbol{(x_{news})}$: Final list of filtered headlines from the aggregation pipeline. The non-finance related headlines were filtered out by performing a similarity search with SBERT model, "all-MiniLM-L6-v2" \cite{reimers2019sentencebert}.  Each headline was compared to a set of artificially generated financial headlines generated by GPT-4, with the prompt \textit{"Generate 20 financial news headlines"}. Headlines with a similarity score below 0.2, were excluded from the dataset. 
To respect the prompting `context length' of LLMs, in instances where the prompt exceeded a length of 3000 words, a further refinement process was employed. This process involved the elimination of words with a tf-idf \cite{tfidf} score below 0.2 and truncating the prompt to a maximum of 3000 words.

It is also important to note that the dataset does not encompass all calendar dates within the specified time range. This limitation emanates from the trading calendar days, and absence of relevant financial news headlines for certain dates.

\paragraph{Label} $\boldsymbol{(x_{r})}$: The label is determined by the percentage change in closing prices from one day to the next, as defined in equation \ref{eq:PCT}. This percentage change is categorized into three labels: \{Rise, Fall, Neutral\}, based on the thresholds specified in equation \ref{eq:label}.

\begin{equation}
    PCT_{\text{change}} = \left( \frac{\text{Closing Price}_{t} - \text{Closing Price}_{t-1}}{\text{Closing Price}_{t-1}} \right) \times 100\%
\label{eq:PCT}
\end{equation}

\begin{equation}
    x_r = 
    \begin{cases}
        \text{Fall} & \text{if } PCT_{\text{change}} < -0.5\% \\
        \text{Neutral} & \text{if } -0.5\% \leq PCT_{\text{change}} \leq 0.5\% \\
        \text{Rise} & \text{if } PCT_{\text{change}} > 0.5\%
    \end{cases}
\label{eq:label}
\end{equation}

\subsection{NIFTY-RL: Preferences Dataset}
The preference dataset is a variation of the fine-tuning dataset and it is designed for alignment training of LLMs using reward model. In NIFTY-RL, labels are omitted and replaced with chosen and rejected results. The chosen result is a label corresponding to a rise, a fall or neutral movement in the stock market and is equivalent to the response in NIFTY-LM. The rejected result is a random label not equal to the chosen label.

\begin{itemize}
    \item \textbf{Metadata}: Includes dates and data identifiers.
    \item \textbf{Prompt ($x_p$)}: Includes an LLM instruction ($x_{question}$), preceding market data ($x_{context}$), and relevant news headlines ($x_{news}$).
    \item \textbf{Chosen Result}: A qualitative movement label ($x_r$) from $\{Rise, Fall, Neutral\}$ indicating the predicted market trend.
    \item \textbf{Rejected Result}: A label ($\overline{x}_r$) randomly selected from $\{Rise, Fall, Neutral, Surrender\} \setminus \{x_r\}$, representing an incorrect market prediction.
    
\end{itemize}





\section{Usage and Applications}\label{sec:applications}
Here we discuss some plausible applications and usage of the contributed dataset. In the following section~\S\ref{sec:experiments}, we provide a few examples of such applications and numerical results as baseline.

\paragraph{Stock Movement (SM) Task} Training and testing of (LLM) experts for the financial market (stock) price movement classification and forecasting may be the foremost and straight-forward application of NIFTY.

\paragraph{Reward based alignment of Language Models}
Tuning pretrained LMs using reward feedback and RL enables remarkable capabilities of current chat-bots and assistants to follow instructions. The RLHF pipeline~\cite{rlhf_ziegler2019fine, rlhf_stiennon2020learning,rlhf_instructgpt_ouyang2022training} is a well-formulated approach in the NLP domain. While variants to RLHF have been proposed~\cite{dpo_rafailov2023}, we discuss only the popular RLHF pipeline for our purposes here. At a high-level, the RLHF pipeline starts with fine-tuning a pre-trained LM in supervised manner (typically with the same LM objective, but on new, high-quality domain-specific data) to obtain $\pi^{SFT}$, then training a reward model $f^{RM}_\theta$ that, once trained, is able to evaluate (usually pairs of) LM generated prompt ($x_p$) completions: $ (\hat{x}^1_r, \hat{x}^2_r) \sim \pi^{SFT}(x_p) $ and provide scalar reward $f^{RM}_\theta(\hat{x}_r) \rightarrow r \in \mathbb{R}$. A human labelled preferences dataset is typically used for the reward model training using MLE objective. In the final step, the domain fine-tuned LM, and the trained reward model is used to fine-tune an aligned policy using RL (e.g. PPO~\cite{ppo_schulman2017proximal}) where $\pi^{SFT}$ acts as the reference based policy: $\pi^{ref}$. PPO uses the base, reference model to impose a KL-divergence penalty during RL fine-tuning using reward feedback to ensure the fine-tuned model does not deviate or diverge too far away from the base policy and preventing unwanted scenarios like mode-collapse to high-reward answers. 
The \texttt{NIFTY-RL} dataset was especially formatted with rejection-sampling labels to facilitate such LLM alignment using desired techniques.

\paragraph{Role of Embeddings in Information Acquisition}
The significant amount of textual data with semantic and temporal connections allow the dataset to be used for a plethora of NLP research topics including answering questions about LM embeddings like: `\textit{do larger models produce richer embeddings}' in terms of information gain? We detail an example experiment design and findings in the ~\S\ref{sec:experiments}.

\paragraph{Regime-Switching in Finance}
The NIFTY dataset enables the SoTA LLM capabilities for advancing research in the challenging financial regime-switching domain.
In empirical finance literature, regime switching processes are modeled as \textit{Markovian Switching Models}, introduced by the seminal work of Hamliton~\cite{markovswitching_hamilton1989new}, in the 1990s. The canonical regime switching problem can be presented by letting $o_t$ be an outcome variable for a market process, which recurrently depends on its own past history, $y_{t-1}$, $\varepsilon_t$ representing random shocks and (for ML/RL community, a conveniently termed) $s_t \in \{0,1,...,k\}$ a discrete random variable modeling some underlying \textit{regime process} at time, $t$. Then regimes affect the intercept(mean), $\mu_{s_t}$, auto-correlation, $\phi_{s_t}$, and volatility, $\sigma_{s_t}$, of the process~\cite{regime_hamilton2010regime}:

\begin{equation}\label{eqn:fin-regime-switching}
    o_t = \mu_{s_t} + \phi_{s_t} o_{t-1} + \sigma_{s_t} \varepsilon_t, 
                                                        \quad \varepsilon_t \sim \operatorname{iid}(0, 1).
\end{equation}

Enthusiastic readers are encouraged to read \cite{regime_guidolin2011markov, markov_switching_hamilton1990analysis, regime_hamilton2010regime} for a detailed overview of Markovian switching models. For a comprehensive appreciation and answer to `\textit{why regime adaptation is important}?', we highly encourage reading~\cite{regime_ang2012regime, regime_guidolin2008size}. Modern deep learning based techniques essentially subsume and skip the problem of regime classification as an intermediary step to some means (like market prediction), and allow the distributional latent embeddings to encapsulate the true regime state from some input data (as a belief $b$ encoding from POMDP formulation). 


\paragraph{Problem Formulation: Market Movement as a POMDP}
Aligned with the regime switching formulation, we model the task of market movement direction as a POMDP problem. We detail pertaining canonical definitions and terminology in the Appendix~\S\ref{app:ss:definitions}. 
Here, we decompose the POMDP problem as an \textbf{MDP over belief states}~\citet{kaelbling1998planning}. Thus, a policy's belief state at time $t$, $b_t$ can be seen as a sufficient statistic of the history $h_t$ towards deciding optimal actions. 

Going forward, observation at time $t$, $o_t$, will be referred to as a LM query, $x_q$ comprised of a prompt $x_{p_t}$ and action prediction label from previous time step: $\hat{x}_{r_{t-1}}$ (Fig.~\ref{fig:nifty-query}).

\section{Experiments }\label{sec:experiments}
Here we present preliminary results as baselines for demonstrating some of the usage and applications of the dataset alluded to in~\S\ref{sec:applications}.


\subsection{Stock Movement (SM) Task}
We show an example of SM classification task using the popular Llama family of LLMs. We explored some variants of each of the presented models by supervised fine-tuning (SFT) of the base language model using NIFTY and three other SM datasets from the recently released Flare Stock Movemente Dataset that  standardizes various existing financial domain evaluation tasks (like sentiment analysis, headlines classification, NER, etc.) using consistent LM queries $x_q$. The benchmark uses the widely adopted \texttt{LM-Eval} LLM evaluation harness~\cite{eval-harness}. The three SM task datasets are: the \textbf{CIKM} datset~\cite{fin-dataset_cikm_wu2018hybrid}, \textbf{StockNet ACL}~\cite{fin-dataset_acl18_stocknet_xu2018stock}, and \textbf{BigData22}~\citep{fin-dataset_bigdata22_soun2022accurate}. Table~\ref{table:flare_sm_datasets} shows their statistics. Full benchmark details is in the appendix~\S\ref{app:flare_datasets}. 

\begin{table}[h]
\caption{Summary of Flare stock price movement datasets.}
\label{table:flare_sm_datasets}
\small
\centering
\setlength{\tabcolsep}{3pt} 
\renewcommand{\arraystretch}{1.2}
\begin{tabular}{l|c|c|c|c|c}
    \hline 
    Data                         & Stocks   & Tweets  & Days & Start Date & End Date \\
    \hline 
    \href{https://github.com/yumoxu/stocknet-dataset}{ACL18}
     & 87       & 106,271 & 696 & 2014-01-02 & 2015-12-30     \\
    \href{https://github.com/stocktweet/stock-tweet}{BigData22} & 50       & 272,762 & 362 & 2019-07-05 & 2020-06-30 \\
    \href{https://github.com/wuhuizhe/CHRNN}{CIKM18}    & 38       & 955,788 & 352 & 2017-01-03 & 2017-12-28    \\
    \hline
\end{tabular}
\end{table}


\begin{table}[h]
\centering
\captionsetup{skip=5pt}
\caption{Performance of a single baseline expert \textbf{Llama-2-7b-chat} with 4 variants (LoRA SFT adapters) on the NIFTY Stock Price Movement Prediction Task (\textit{test split}).}
\label{tab:Llama-2-7b-chat-hf_nifty-combined-results}
\resizebox{\columnwidth}{!}{%
    \renewcommand{\arraystretch}{1.2} 
    \begin{tabular}{@{}lccccc@{}}
    \toprule
    & Base Expert & \multicolumn{4}{c}{Expert Variants (SFT w/ LORA Adapters)} \\
    \cmidrule(lr){2-2} 
    \cmidrule(lr){3-6}
    Metrics $\uparrow$ & \begin{tabular}[c]{@{}c@{}}Llama-2-7b\\chat\end{tabular} & \begin{tabular}[c]{@{}c@{}}Llama-2-7b\\chat + nifty\end{tabular} & \begin{tabular}[c]{@{}c@{}}Llama-2-7b\\chat + acl18\end{tabular} & \begin{tabular}[c]{@{}c@{}}Llama-2-7b\\chat + bigdata22\end{tabular} & \begin{tabular}[c]{@{}c@{}}Llama-2-7b\\chat + cikm18\end{tabular} \\
    \midrule
    F1 Score & $0.24$ & $0.21$ & $0.23$ & $0.28$ & $0.28$ \\
    Accuracy & $0.26$ & $\textbf{0.45}$ & $0.25$ & $0.29$ & $0.28$ \\
    Precision & $0.32$ & $0.15$ & $0.31$ & $0.29$ & $0.29$ \\
    Recall & $0.33$ & $\textbf{0.33}$ & $0.29$ & $0.28$ & $0.28$ \\
    \bottomrule
    \end{tabular}%
}
\end{table}

\begin{table}[h]
\centering
\captionsetup{skip=5pt}
\caption{Performance of a single baseline expert \textbf{Meta-Llama-3-8B-Instruct} with 4 variants (LoRA SFT adapters) on the NIFTY Stock Price Movement Prediction Task (\textit{test split}).}
\label{tab:Meta-Llama-3-8B-Instruct_nifty-combined-results}
\resizebox{\columnwidth}{!}{%
    \renewcommand{\arraystretch}{1.2} 
    \begin{tabular}{@{}lccccc@{}}
    \toprule
    & Base Expert & \multicolumn{4}{c}{Expert Variants (SFT w/ LORA Adapters)} \\
    \cmidrule(lr){2-2} 
    \cmidrule(lr){3-6}
    Metrics $\uparrow$ & \begin{tabular}[c]{@{}c@{}}Llama-3-8B\\Instruct\end{tabular} & \begin{tabular}[c]{@{}c@{}}Llama-3-8B\\Instruct + nifty\end{tabular} & \begin{tabular}[c]{@{}c@{}}Llama-3-8B\\Instruct + acl18\end{tabular} & \begin{tabular}[c]{@{}c@{}}Llama-3-8B\\Instruct + bigdata22\end{tabular} & \begin{tabular}[c]{@{}c@{}}Llama-3-8B\\Instruct + cikm18\end{tabular} \\
    \midrule
    F1 Score & $0.29$ & $\textbf{0.30}$ & $0.20$ & $0.26$ & $0.25$ \\
    Accuracy & $0.38$ & $\textbf{0.40}$ & $0.26$ & $0.30$ & $0.29$ \\
    Precision & $0.26$ & $0.27$ & $0.25$ & $\textbf{0.29}$ & $0.28$ \\
    Recall & $0.33$ & $\textbf{0.35}$ & $0.29$ & $0.30$ & $0.29$ \\
    \bottomrule
    \end{tabular}%
}
\end{table}


\subsection{Role of LM embeddings in information acquisition}

Here, we extend the discussion of model scalability and its implications on semantic clustering. Specifically, we explore the question: `\textit{Do larger models produce richer embeddings}' in terms of information gain using the NIFTY dataset.

Our experiments here, 
detailed in Appendix~\S\ref{app:embeddings}, 
demonstrate the \textbf{Hypothesis}: \textbf{larger models generate more informative embeddings}, which in turn enhance the granularity of semantic clustering. This increased granularity is particularly relevant in the financial domain, where precise interpretation of market-related news can significantly impact predictive accuracy. 

By leveraging higher-dimensional vector spaces provided by these larger models, we observe a clear increase in \textit{information gain} for market movement, location (Fig.~\ref{fig:3x3grid}), and genre tasks. These findings corroborate our hypothesis regarding the critical role of model size in semantic analysis and forecasting in finance. 

\definecolor{darkpurple}{RGB}{51, 0, 102} 
\begin{figure*}[ht]
    \centering
    \begin{subfigure}[b]{0.24\textwidth}
        \includegraphics[width=\textwidth]{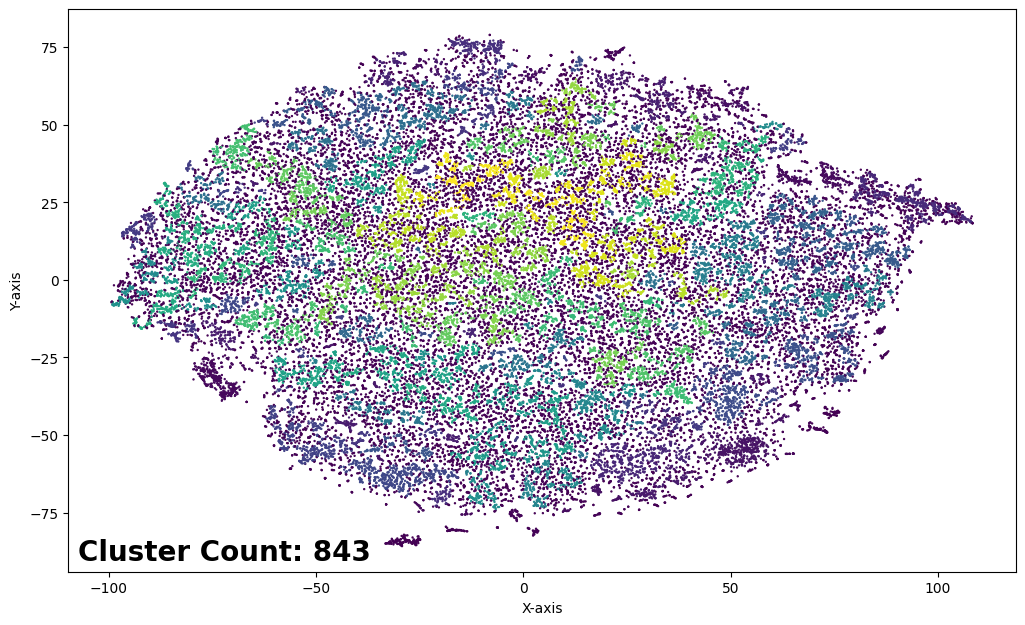}
        \caption{GPT2-SMALL}
        \label{fig:small_location}
    \end{subfigure}
    \hfill
    \begin{subfigure}[b]{0.24\textwidth}
        \includegraphics[width=\textwidth]{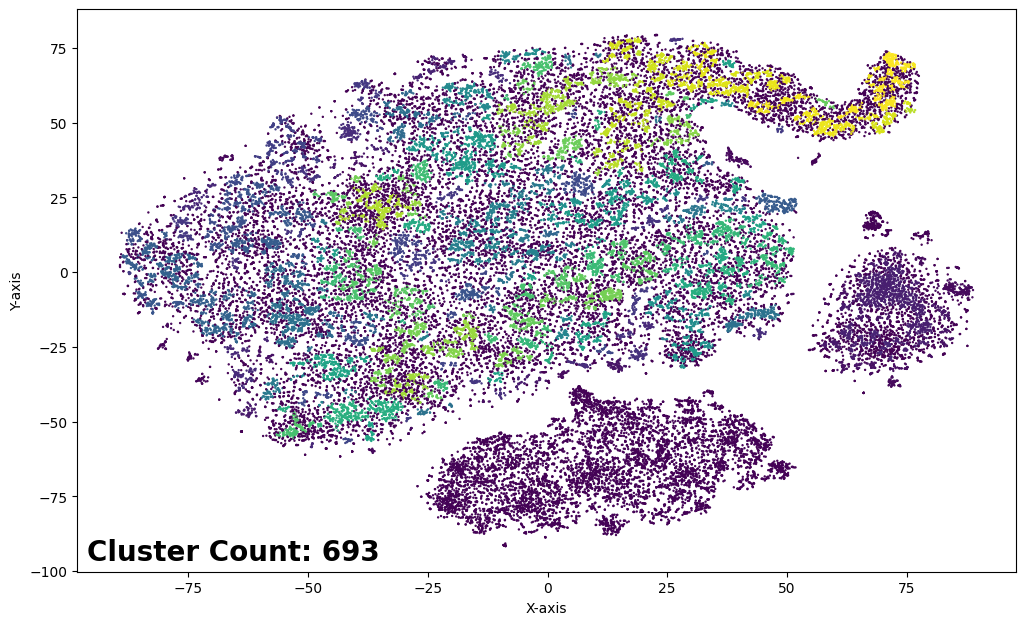}
        \caption{GPT2-MEDIUM}
        \label{fig:medium_location}
    \end{subfigure}
    \hfill
    \begin{subfigure}[b]{0.24\textwidth}
        \includegraphics[width=\textwidth]{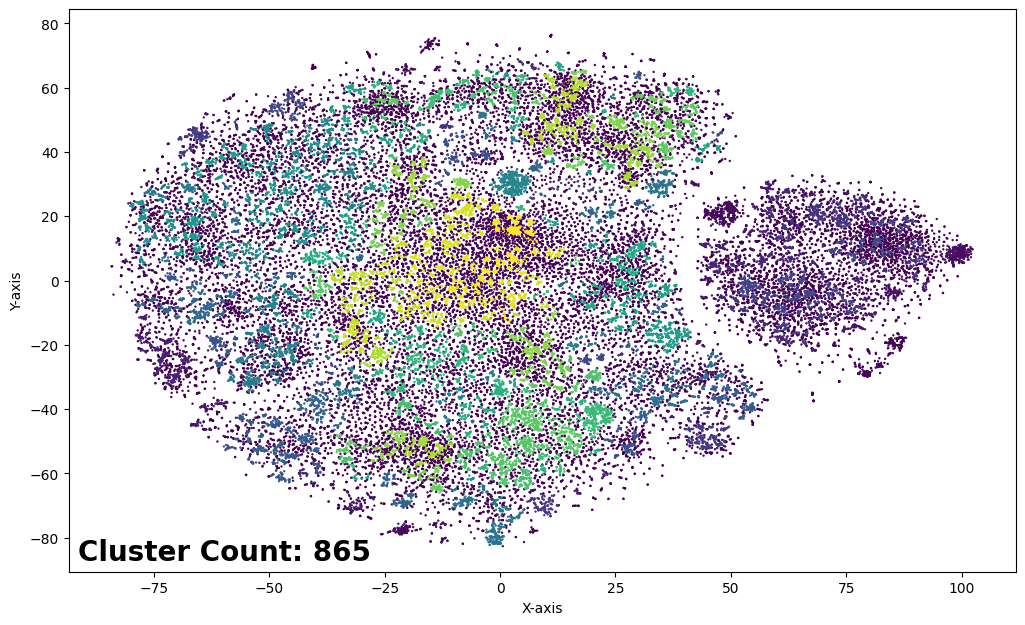}
        \caption{GPT2-Large}
        \label{fig:large_location}
    \end{subfigure}
    \hfill
    \begin{subfigure}[b]{0.24\textwidth}
        \includegraphics[width=\textwidth]{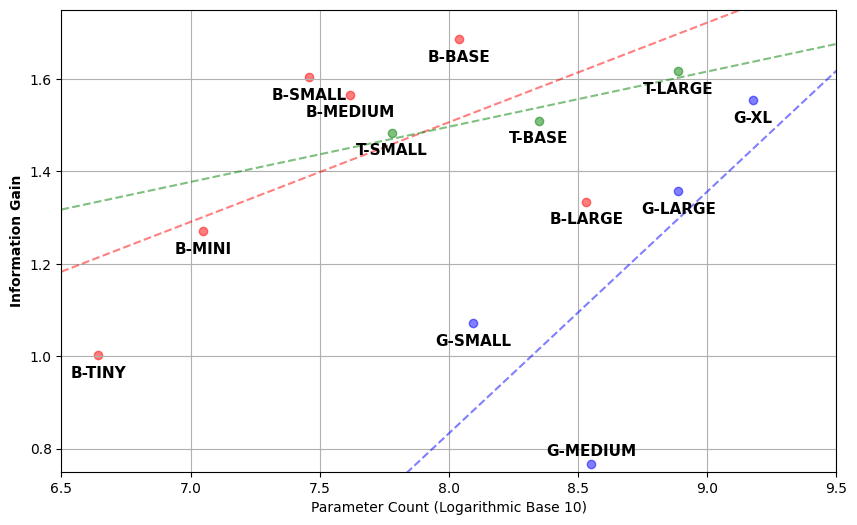}
        \caption{Location Information Gain}
        \label{fig:location_results}
    \end{subfigure}
    
    \caption{\textbf{(a-c)}: Visualizations of 2D t-SNE projections of embedded clusters (using HDBSCAN with minimum cluster size of 10) for models GPT2-[SMALL, MEDIUM, LARGE]. Each datapoint is an embedding of a news headline with a location tag in [U.S, Europe, Asia, Middle East, Latin America]. Each colour is associated with a cluster of headlines. The background purple hue are datapoints belonging to the \textcolor{darkpurple}{outlier} cluster. 
    \textbf{(d)}: \textit{Information gain} added when clustering model embeddings together on the headline location task. Information gain increases with number of model parameters. Pattern persists across model architectures: GPT2 models are shown in \textcolor{blue}{blue}, BERT models in \textcolor{red}{red}, and T5 models in \textcolor{green}{green}.}
    \label{fig:3x3grid}
\end{figure*}

\newpage \clearpage

\bibliographystyle{plain}
\bibliography{refs/main, refs/rl, refs/irl, 
                refs/qfin-datasets, 
                refs/qfin-pm,
                refs/benchmarks,
                refs/llms,
                refs/kalman-filters,
                refs/nvinden-ref-search}

\newpage
\appendixpage
\DoToC
\appendix

\section{Definitions and Terminology}\label{app:ss:definitions}

\paragraph{Markov Decision Process (MDP)} An MDP is defined by a tuple $(S, A, T, R, \gamma, p_0)$ where $S$ is a set of states (state space), $A$ is a set of actions, $T: S\times A \to \Pi(S)$ is the transition function, $R: S \to \mathbb{R}$ is the reward function, $\gamma \in [0, 1]$ is the discount factor, and 
$p_0: S \to [0, 1]$ 
is the distribution over initial states. A policy over an MDP is a function $\pi: S \to \Pi(A)$, and is optimal if it maximizes the expected discounted sum of rewards.

\begin{equation}
\mathcal{L} = \mathbb{E}_{\pi, T}\left(\sum_{s_i \in \tau} \gamma^i R(s_i)\right),    
\end{equation}
where $\tau = (s_0, a_0, \dots, s_T)$ is a trajectory. 

\paragraph{Partially Observable Markov Decision Process (POMDP)} 
A POMDP is a generalisation of an MDP defined by the tuple $(S, A, T, O, \omega, R, \gamma, p_0)$ where $O$ is a set of observations and $\omega: S \to \Pi(O)$ is the \textit{observation function}. An agent in a POMDP thus only receives an observation (i.e., partial information about the state) rather than the actual state of the environment. Therefore, policies on POMDPs act based on the history of observations received and actions taken at timestep $t$.

\paragraph{Belief MDPs}
Since using the complete history is impractical, many algorithms instead use  \textit{belief states} $b: O \to \Pi(S)$, which is a probability distribution over possible states updated at each timestep, given history $h_t$ comprising of previous observations. 
Intuitively, it can be thought of an agent maintaining a `belief' – a probability distribution over what it thinks the true state of the environment might be.

The belief update after taking the action $a\in A$ and receiving observation $o \in O$ is done through the following equation:
\begin{align}
    \label{eqn:belief}
    b_{o}^{a}\left(s^{\prime}\right) &= P\left(s^{\prime} \mid b, a, o\right) \nonumber \\
    &= \frac{\omega\left(s^{\prime}, o\right) \sum_{s} T\left(s, a, s^{\prime}\right) b(s)}{P(o \mid b, a)} \quad \forall s' \in S,
\end{align}
where $P(o \mid b, a)=\sum_{s^{\prime}} \omega\left(s^{\prime}, o\right) \sum_{s} T\left(s, a, s^{\prime}\right) b(s)$.

We can formulate any POMDP problem as an MDP over belief states~\citet{kaelbling1998planning}. Thus, an agent's belief state at time $t$, $b_t$ can be seen as a sufficient statistic of the history $h_t$ towards deciding optimal actions.

\section{Do Larger Models Produce Richer Embeddings?}\label{app:embeddings}

\begin{figure}[ht]
    \centering
    \includegraphics[width=0.8\textwidth]{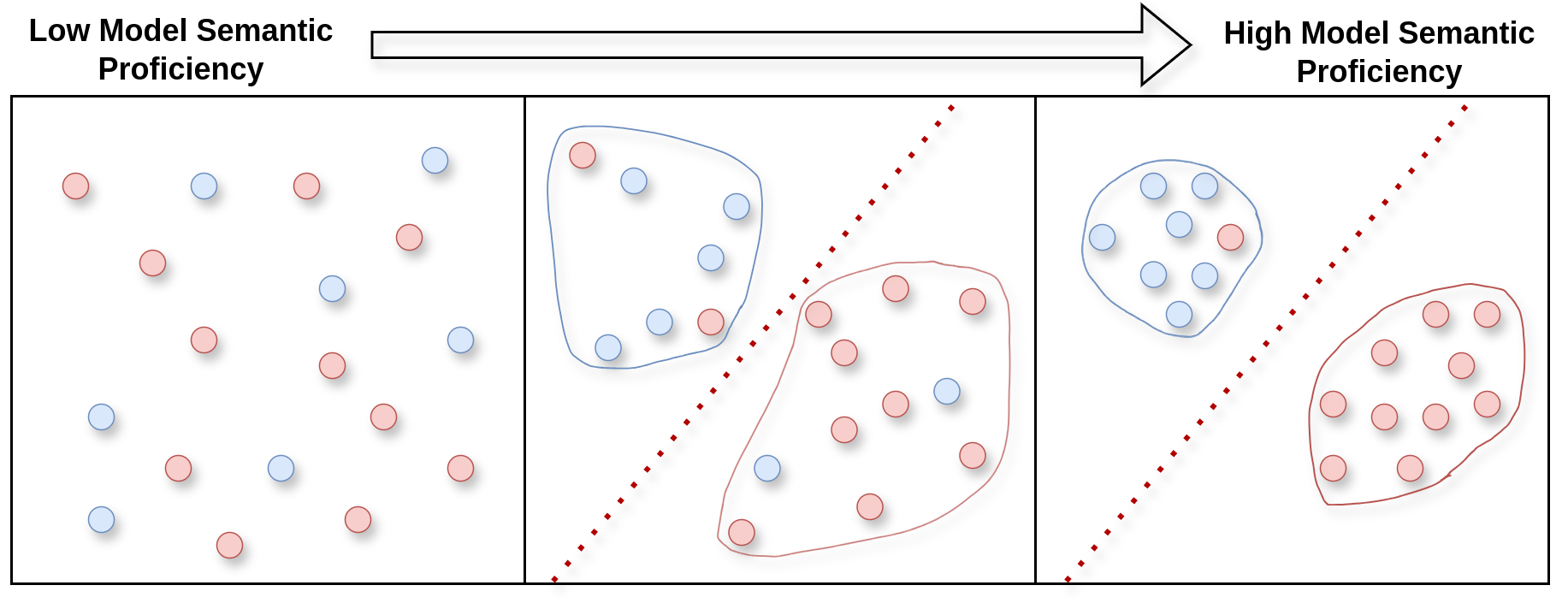}
    \caption{\textbf{Information Gain in Clustered Prompt Embeddings (IG-CluPE):} A novel method of measuring a LLM's ability to capture rich semantic contextualization of a corpus of text prompts with corresponding classifications. Prompt embeddings are extracted from outputs of the last-hidden-layer of transformer models to create an embedding space optimized for linear separability of points from each class. The effectiveness of a model's ability to group points with similar features together is measured through t-SNE clustering and information gain.}
    \label{fig:information_gain_cluster}
\end{figure}

In this section, we do an analysis of prompt embeddings to provide evidence for the efficacy of LLM alignment finetuning, and the information density of NIFTY over other Flare stock movement datasets. When processing prompts, transformer models like LLaMA-2 \cite{llama2-touvron2023llama} produce large-dimensional vectors that capture the structure and semantic features. Consequentially, prompt embeddings localized in a group contain more similar semantic features than those of sentences of further distance in the embedding space \cite{wieting-etal-2017-learning}. Specifically, we investigate to what degree a model's embedding is able to group prompts identical market movement directions together. To do this we use t-SNE \cite{maaten2008visualizing} to generate embeddings for all NIFTY, ACL18, BigData22, and CIKM18 prompts, we measure the Information Gained (IG) after clustering.

\paragraph{IG-CluPE: Information Gain in Clustered Prompt Embeddings}
Generating rich embeddings from large langugage model prompts has been a key research topic for a number of years now. A properly trained LLM can produce an embedding space that captures deep prompt-wise semantic relationships. Using pretrained transformer \cite{vaswani2017attention} based architectures such as GPT \cite{gpt3-brown2020language}, T5 \cite{2020t5}, BERT \cite{kenton2019bert}, and LLaMA-2 \cite{llama2-touvron2023llama} have a proven ability to capture semantic structures of prompts in richest ways possible. With IG-CluPE, we propose a method that uses prompt embeddings from LLaMA-2's last hidden layer, and we use information gain in prompt clustering to measure the information density of the produced embedding space in order to capture the power of our models and to measure information richness of a suite of financial market movement datasets.

Generating information gain of an embedded space from IG-CluPE is outlined in these steps: 

\paragraph{1. Embedding Generation:} We feed through each tokenized prompt ($x_p$) through our LLaMA-2 model, extracting and saving the outputs from the final hidden layer of the transformer as prompt embeddings.

\paragraph{2. Prompt Clustering:} Once embeddings are generated for all prompts, we use \textit{t-distributed Stochastic Neighbor Embedding} (t-SNE)~\cite{maaten2008visualizing} to cluster all prompts. For purposes of visualization we also use HDBSCAN~\cite{campello2013density} for creating cluster figures in Cartesian space.

\paragraph{3. Information Gain Measurement:} We measure the information gain of clustering each prompt with equation \ref{eq:informationGain}, where \( L \) is a set of \( M \) tags \( (l_1, l_2, \ldots, l_M) \), \( T \) is a multiset of \( N \) tags such that each element \( t \in T \) is also in \( L \), and \( \{P_1, P_2, \ldots, P_K\} \) is a partition of \( T \) into \( K \) clusters.

\begin{align}
p(l, T) &= \frac{|\{ i \in T : \text{label of } i = l \}|}{|T|} \label{eq:probability} \\
H(T) &= -\sum_{l \in L} p(l, T) \log_2 p(l, T) \label{eq:Entropy} \\
H_{C}(P) &= \sum_{k=1}^{K} \frac{|P_k|}{|T|} H(P_k) \label{eq:clusteredEntropy} \\
IG &= H_{C}(P) - H(T) \label{eq:informationGain}
\end{align}

\paragraph{Intuition and reasoning} The intuition behind using a last-hidden-layer embedding clustering-based approach to measure information richness is rooted in the optimization processes of classification models. By analyzing the embeddings from the final hidden layer of a neural network, we can assess how well the model captures and discriminates between different classes of data. Clustering these embeddings allow us to both qualitatively and quantitatively evaluate the separability and density of the data representations, reflecting the model's ability to generalize and its sensitivity to various financial features. This approach not only offers insights into the model's internal representations but also allows us to generalize which datapoints share pertinent features to stock movement prediction.

By using the last-hidden-layer of a transformer architecture, preceding a single-layer fully connected neural module, we ensure that during training the model is optimized for linear separability of the last-hidden-layer embedding space. Therefore, we pose that in an optimized model a single data point has an increased probability of being surrounded by data points of the same class, as compared to a worse performing model. We then borrow a technique of measuring the degree of cluster homogeneity through information gain in decision tree optimization, described in equations \ref{eq:probability} - \ref{eq:informationGain}.

\paragraph{Related Works with Last Layer Embeddings} Generating prompt embeddings with LLMs is a rich and vibrant field of study, owing to their usefulness in knowledge representation in a field dominated by ``black box'' algorithms. Examples like Jiang et al. (2023) \cite{jiang2023scaling} enhancing sentence embeddings through in-context learning, Kervadec et al. (2023) \cite{kervadec2023unnatural} analyzing responses to machine-generated prompts, reveal significant differences in model responses and network processing pathways.

\paragraph{Advantages of IG-CLuPE over classification accuracies} Although we state that a model's IG-CluPE score is proportional of the model's ability to perform a downstream classification task it was optimized for, we find that using \textbf{IG-CluPE has a set of marked advantages} over only using classification accuracy for model evaluation. Clustering prompt allows us to better interpret model decision, and lets us view which prompt features the model finds useful in prompt classification. Whereas viewing model performance solely through the lens of classification accuracy groups each prompt into one of N categories, IG-CluPE allows us to peer into the prompt space and visualize how the model groups similar points. We can look at false negatives and see which other prompts are closest to that prompt in the embedding space.

Model information density analysis through IG-CluPE is additionally insightful in the domain of comparing the efficacy of similar LLM models, or when comparing the information density of similar datasets. A IG-CluPE can guide model design by peering into the inner workings of the model and identifying weak points. For example, in the context of semantic classification, if a model predominantly groups prompts of classes \textit{HAPPY} and \textit{EUPHORIC} together, we could tweak training methodology to include more cases of these classes in the dataset. Then another embedding space can be created, and results compared. Additionally, we can also look at how information density in the context of identically trained models embedding prompts of similar datasets. A dataset with a modified/additional set of features can guide the models ability to correctly classify text phrases. A clustered embedding space for each dataset can highlight how our model utilize changes in the feature set. 

In this section, we test both of these cases by using IG-CluPE to measure information richness of models with three distinct architecture types (encoder-only, decoder-only and encoder-decoder) with increasing sizes (by model parameters).

\subsection{Experiments}

Testing if larger models create richer embeddings is predicated on a model's ability to group datapoints with similar features together. The features we measure are realized in \textbf{three tasks}: \textit{market movement, location}, and \textit{genre}. Each headline in the NIFTY dataset contains a single ``Tag'' that acts as a label for the category to which the headline belongs. For each task, we subsample the NIFTY, taking only task-specific tags and omitting all other rows. The tags used in the \textit{location} and \textit{genre} tasks are designed to be mutually exclusive, so a data point cannot correctly belong to two clusters. A well-performing model will create homogeneous clusters, consisting of headlines with the same tag. 

For the market movement task, we are interested in measuring how an LLM's semantic perception of a news headline can be indicative of market movement, so we only include tags relating to markets and finance. Further, in this task, we are not interested in clusters with the homogeneous tags, but instead we measure whether headlines in a cluster are indicative of homogeneous market movement. A well performing model clusters points with similar direction and magnitude of market movement from the date that each headline was published. Datasets $NIFTY_{L}$, $NIFTY_{G}$, and $NIFTY_{MM}$ are created as subsets of NIFTY with only their corresponding tags. The tags used are shown in Table~\ref{table:task_summary}.

\begin{table}[ht]
\centering
\caption{Summary of Tasks and Their Characteristics}
\label{table:task_summary}
\begin{tabular}{lcc}
\toprule
Task & Tags & Headline Count \\ 
\midrule
Market Movement & Finance, Business, Markets, Earnings & 69,068 \\
\midrule
Location & U.S, Europe, Asia, Middle East, Latin America & 49,446 \\
\midrule
Genre & Politics, World News, Business and Economy, & 382,698 \\
& Science and Environment, Health, Entertainment, \\
& Sports, Opinion and Editorial, Human Interest \\
\bottomrule
\end{tabular}
\end{table}

For each model architecture, we test multiple sizes of pretrained models, each with a different number of parameters. Each model was tested using Huggingface's transformer package \cite{huggingface_transformers}, with the exception of the OPENAI-ADA2, OPENAI-SMALL, and the OPENAI-LARGE models, whose embeddings were gathered using OpenAI's API \cite{OpenAI_API}. Parameter counts have not been disclosed for any of their embedding models, however OpenAI have noted that OPENAI-SMALL is a larger model than OPENAI-LARGE. For the T5 models, we used the small, base, and large models; and for the BERT models we used the tiny, mini, small, medium and base models. Parameter counts for each public model are available in Table~\ref{table:embedding_results}. Model's GPT2, T5, and BERT were chosen to include a decoder-only, encoder-decoder, and encoder-only model respectively.

For each model, we generated embeddings for each headline in $NIFTY_{MM}$, $NIFTY_{L}$, and $NIFTY_{G}$ datasets. Each model inputs a tokenized headline and outputs an embeddings (model embeddings are shown in ~\ref{table:embedding_results}). In order to better visualize each embedding space, we used the \textit{t-distributed Stochastic Neighbor Embedding} (t-SNE)~\cite{maaten2008visualizing} algorithm in order to reduce the dimensions of each embedding into 2 dimensions that are then plotted. t-SNE was chosen as its density-based approach outperformed principle component analysis (PCA) and uniform manifold approximation and projection (UMAP)~\cite{pearson1901liii, 2018arXivUMAP} in putting headlines into discrete clusters.

After the dimensionality of each embedding is reduced to 2 with t-SNE, we use HDBSCAN~\cite{campello2013density} to cluster our set of datapoints into discrete clusters. We require a minimum cluster size of 10 points. Datapoints that do not fit into a cluster are marked as outliers and put into their own ``outlier" cluster.

To quantify the information gain achieved through clustering, we initially computed the entropy of the unclustered multiset of tags in $NIFTY_{L}$, denoted $T_{L}$. The entropy for the base tags, $H(T_L)$, was calculated using the equation \ref{eq:Entropy}. Following clustering with HDBSCAN, we computed the total entropy of the set of clusters P, $H_C(P_T)$, using equation \ref{eq:clusteredEntropy}. Information gain associated with the clustering of location tags in described in equation \ref{eq:informationGain}, and produced $IG_{L} = H_{C}(P_T) - H(T_L)$ \cite{shannon1948mathematical}. This process is repeated for the genre tasks, using dataset $NIFTY_{G}$.

\begin{align}
p(l, T) &= \frac{|\{ i \in T : \text{label of } i = l \}|}{|T|} \label{probability} \\
H(T) &= -\sum_{l \in L} p(l, T) \log_2 p(l, T) \label{eq:Entropy} \\
H_{C}(P) &= \sum_{k=1}^{K} \frac{|P_k|}{|T|} H(P_k) \label{eq:clusteredEntropy} \\
IG &= H_{C}(P) - H(T) \label{eq:informationGain}
\end{align}

where \( L \) is a set of \( M \) tags \( (l_1, l_2, \ldots, l_M) \), \( T \) is a multiset of \( N \) tags such that each element \( t \in T \) is also in \( L \), and \( \{P_1, P_2, \ldots, P_K\} \) is a partition of \( T \) into \( K \) clusters.

For the market movement task, each headline is associated with a percent daily change in market value. Given its continuous nature, we adopted a variance-based approach as an alternative to information gain \cite{hastie2009elements}. The initial variance, $\sigma^2(T)$ (equation\ref{eq:baseVariance}), was calculated across the embeddings before clustering. Post-clustering, the variance within each cluster, $\sigma^2(P_k)$, was computed, and a weighted sum of these variances provided the overall variance after clustering, $\sigma^2_{\text{C}}(P)$ (equation \ref{eq:weightedClusterVariance}). The reduction in variance, is denoted $RV$, and is described in equation \ref{eq:infoGainVariance}. 

\begin{align}
\sigma^2(T) &= \text{Var}(T) \label{eq:baseVariance} \\
\sigma^2_{\text{C}}(P) &= \sum_{k=1}^{K} \frac{|P_k|}{|T|} \sigma^2(P_k) \label{eq:weightedClusterVariance} \\
RV &= \sigma^2(T) - \sigma^2_{\text{C}}(P) \label{eq:infoGainVariance}
\end{align}

This variance reduction approach aligns with our objective to discern the LLM's capability to semantically cluster financial news in a manner indicative of market movement. A model that is able to cluster headlines with similar percent-changes in market movement, leads to low per-cluster market movement variance, and a higher levels of information gained post-clustering.

\subsection{Results and Main Findings}

\begin{figure}[h]
    \centering
    \begin{subfigure}[b]{0.32\textwidth}
        \includegraphics[width=\textwidth]{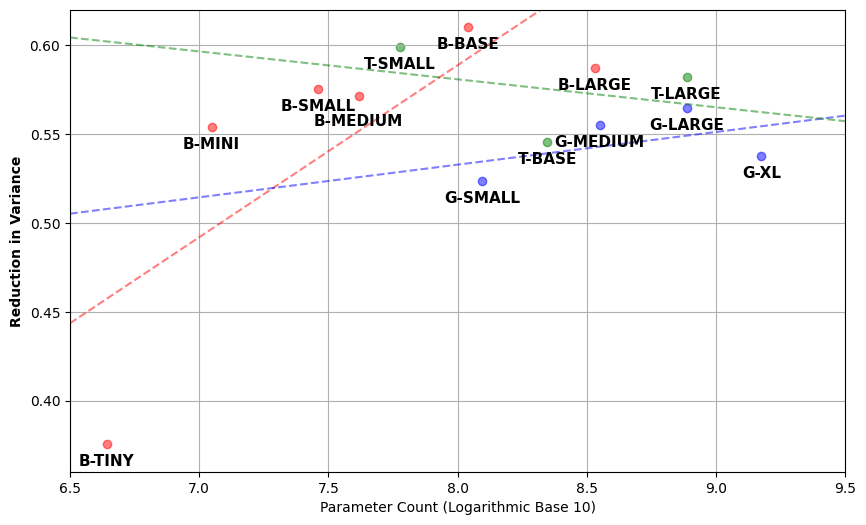}
        \caption{Market Movement Task}
        \label{fig:market_movement_results}
    \end{subfigure}
    \hfill 
    \begin{subfigure}[b]{0.32\textwidth}
        \includegraphics[width=\textwidth]{figs/nvinden_appx/location_vs_IG.png}
        \caption{Location Task}
        \label{fig:location_results_apdx}
    \end{subfigure}
    \hfill 
    \begin{subfigure}[b]{0.32\textwidth}
        \includegraphics[width=\textwidth]{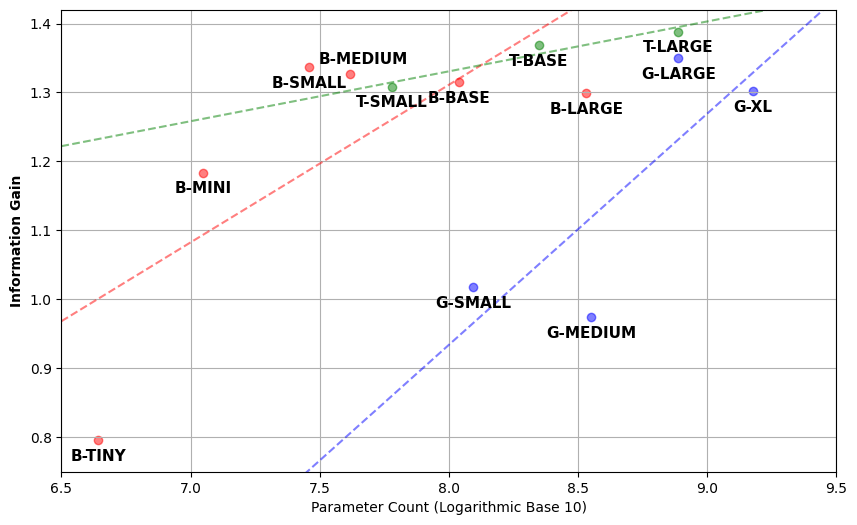}
        \caption{Genre Task}
        \label{fig:genre_results}
    \end{subfigure}
    \caption{Reduction in variance \textbf{(a)} and information gain \textbf{(b-c)} added when clustering model embeddings together on the market movement, location, and genre tasks. Multiple sizes of GPT2 (\textcolor{blue}{blue}), T5 (\textcolor{green}{green}), and BERT (\textcolor{red}{red}) models are plotted with trend line showing increase in parameter count leading to higher clustered reduction in variance and information gain. Strong correlations between parameter count and information gain are shown for all 3 model types in the location and genre tasks. In the market movement task, variance is reduced when parameter counts are increased for the GPT2 and BERT models, but not for T5 models. Although not shown in (a-c), due to having undisclosed parameter counts, OPENAI-LARGE outperformed OPENAI-SMALL in each task. All results are available in Table \ref{table:embedding_results}.}
    \label{fig:datasets}
\end{figure}

Information gain resulted from clustering our list of model's embeddings are summarized in Table~\ref{table:task_summary}, and Figure~\ref{fig:datasets}. Overall, we find that there is a \textbf{strong trend that models with a larger amount of parameters have a higher amount of information gain} in the market movement, location, and genre tasks. This leads credence to imply that larger models have the capability of creating richer embeddings on a plethora of tasks, and using larger models can lead to bigger gains in downstream tasks such as predicting market movement.

Images of subset of model clusters are available in Figure~\ref{fig:3x3grid}.

\begin{table}[ht]
\centering
\caption{Model Performance and Information Gain}
\label{table:embedding_results}
\begin{tabular}{lcccccc}
\toprule
\multirow{2}{*}{Model}  & \multirow{2}{*}{Parameter Count} & \multirow{2}{*}{Embedding Size} & \multicolumn{1}{c}{Reduction of Variance} & \multicolumn{2}{c}{Information Gain} \\ \cline{4-6} 
                        &                                  &                                 & Market Movement & Location     & Genre      \\ 
\midrule
BERT-TINY               & 4M                               & 128                             & 0.38         & 1.00         & 0.80       \\
BERT-MINI               & 11M                              & 256                             & 0.55         & 1.27         & 1.18       \\
BERT-SMALL              & 29M                              & 512                             & 0.58         & 1.60         & \textbf{1.34}       \\
BERT-MEDIUM             & 41M                              & 512                             & 0.57         & 1.57         & 1.33       \\
BERT-BASE               & 109M                             & 768                             & \underline{\textbf{0.61}}        & \textbf{1.68}        & 1.32       \\ 
BERT-LARGE              & 340M                             & 1024                            & 0.59         & 1.33         & 1.30       \\ 
\midrule
T5-SMALL                & 60M                              & 512                             & \textbf{0.60}         & 1.48         & 1.31       \\
T5-BASE                 & 222M                             & 768                             & 0.55         & 1.51         & 1.37       \\
T5-LARGE                & 770M                             & 1024                            & 0.58         & \textbf{1.61}         & \underline{\textbf{1.39}}       \\ 
\midrule
GPT2-SMALL              & 124M                             & 768                             & 0.52         & 1.07         & 1.02       \\
GPT2-MEDIUM             & 355M                             & 1024                            & 0.55         & 0.77         & 0.97       \\
GPT2-LARGE              & 774M                             & 1280                            & \textbf{0.56}         & 1.36         & \textbf{1.35}       \\
GPT2-XL                 & 1.5B                             & 1600                            & 0.53         & \textbf{1.55}         & 1.30       \\
\midrule
OPENAI-SMALL            & -                                & 1536                            & 0.45         & 1.88         & 1.33       \\
OPENAI-LARGE            & -                                & 3072                            & \textbf{0.49}         & \underline{\textbf{1.89}}         & {\textbf{1.36}}       \\
\bottomrule
\end{tabular}
\end{table}

\begin{figure}[h]
    \centering
    \begin{subfigure}[b]{0.3\textwidth}
        \includegraphics[width=\textwidth]{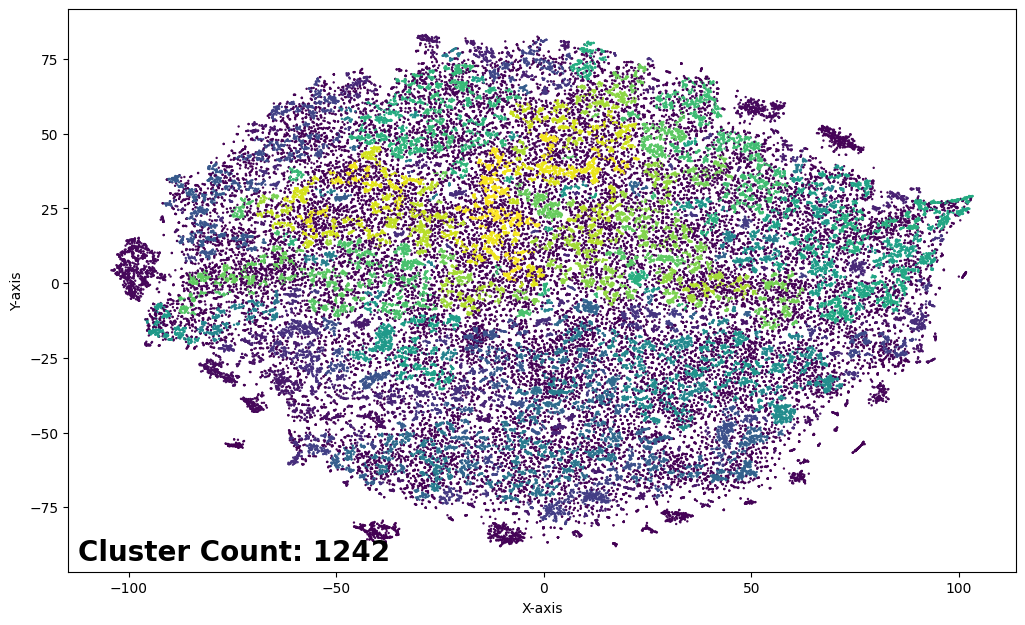}
        \caption{SMALL - Market Movement}
        \label{fig:gpt2_market}
    \end{subfigure}
    \hfill
    \begin{subfigure}[b]{0.3\textwidth}
        \includegraphics[width=\textwidth]{figs/nvinden_appx/small_location.png}
        \caption{SMALL - Location}
        \label{fig:gpt2_location}
    \end{subfigure}
    \hfill
    \begin{subfigure}[b]{0.3\textwidth}
        \includegraphics[width=\textwidth]{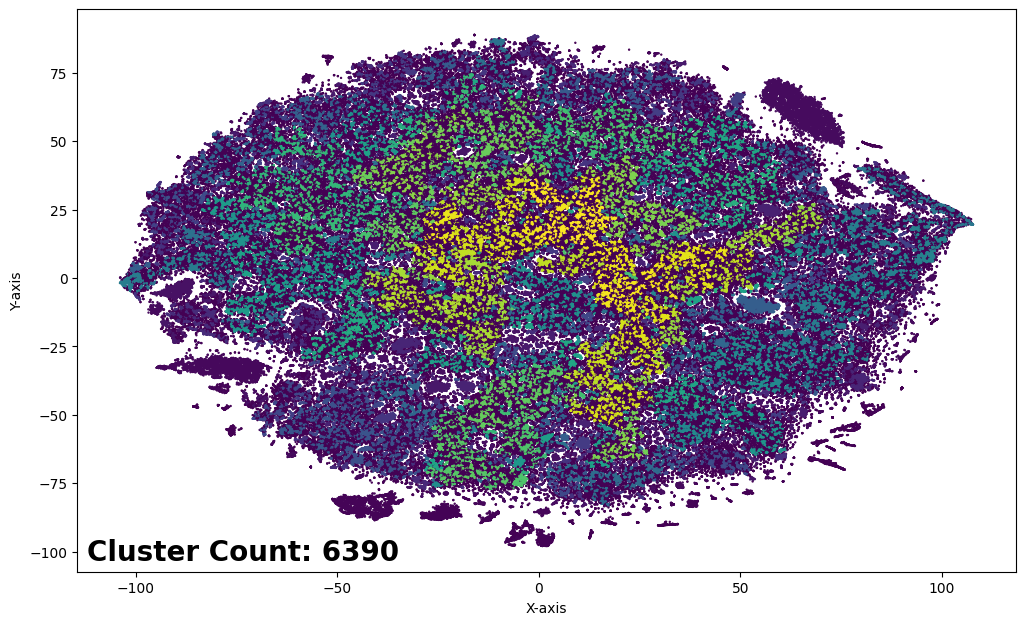}
        \caption{SMALL - Genre}
        \label{fig:gpt2_genre}
    \end{subfigure}

    \begin{subfigure}[b]{0.3\textwidth}
        \includegraphics[width=\textwidth]{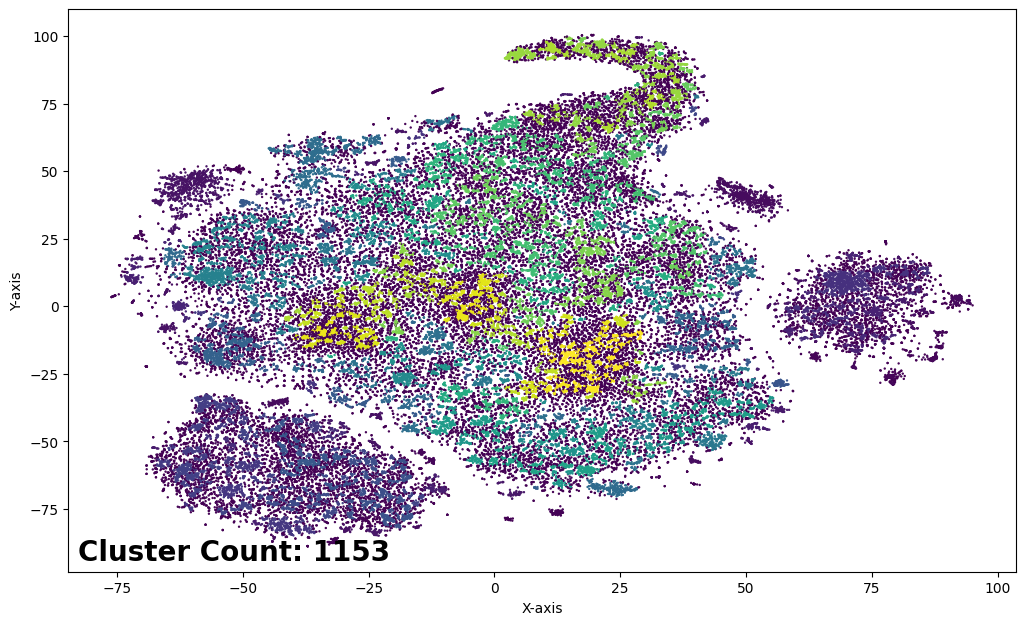}
        \caption{MEDIUM - Market Movement}
        \label{fig:small_market}
    \end{subfigure}
    \hfill
    \begin{subfigure}[b]{0.3\textwidth}
        \includegraphics[width=\textwidth]{figs/nvinden_appx/medium_location.png}
        \caption{MEDIUM - Location}
        \label{fig:small_location_apdx}
    \end{subfigure}
    \hfill
    \begin{subfigure}[b]{0.3\textwidth}
        \includegraphics[width=\textwidth]{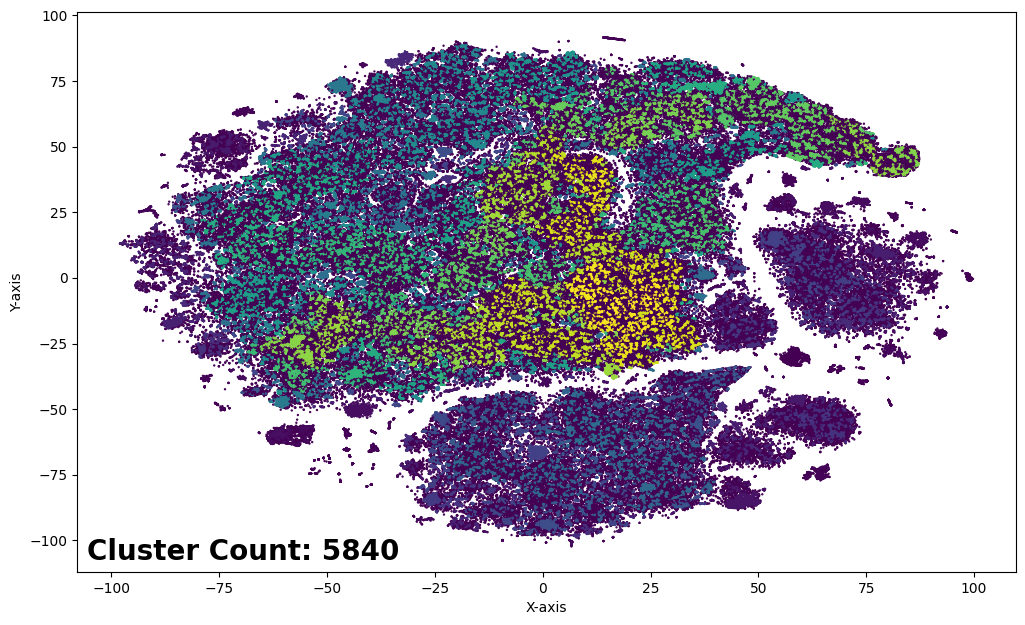}
        \caption{MEDIUM - Genre}
        \label{fig:small_genre}
    \end{subfigure}

    \begin{subfigure}[b]{0.3\textwidth}
        \includegraphics[width=\textwidth]{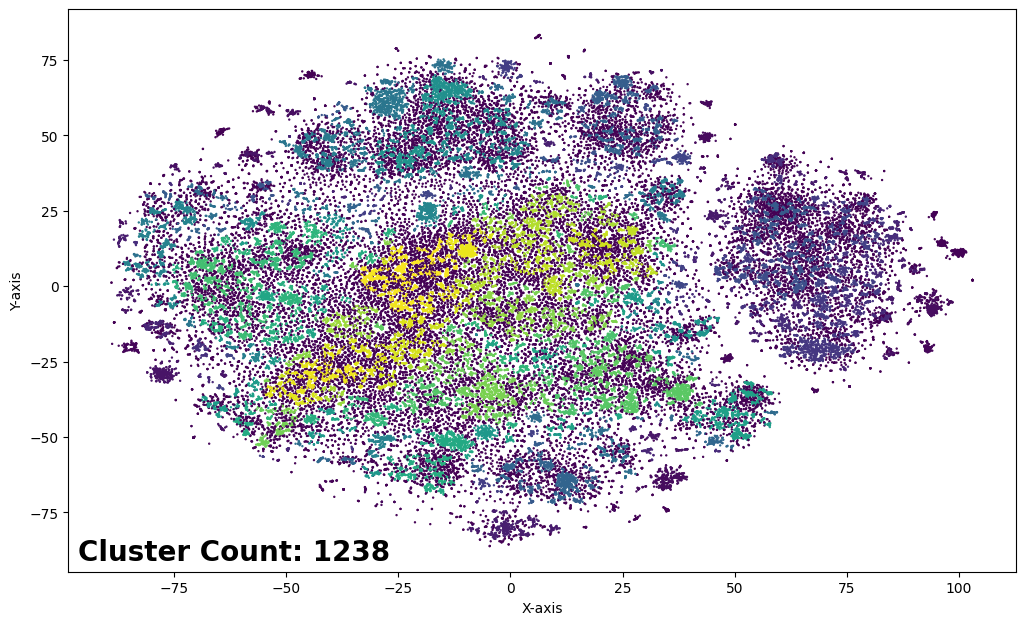}
        \caption{LARGE - Market Movement}
        \label{fig:large_market}
    \end{subfigure}
    \hfill
    \begin{subfigure}[b]{0.3\textwidth}
        \includegraphics[width=\textwidth]{figs/nvinden_appx/large_location.png}
        \caption{LARGE - Location}
        \label{fig:large_location_apdx}
    \end{subfigure}
    \hfill
    \begin{subfigure}[b]{0.3\textwidth}
        \includegraphics[width=\textwidth]{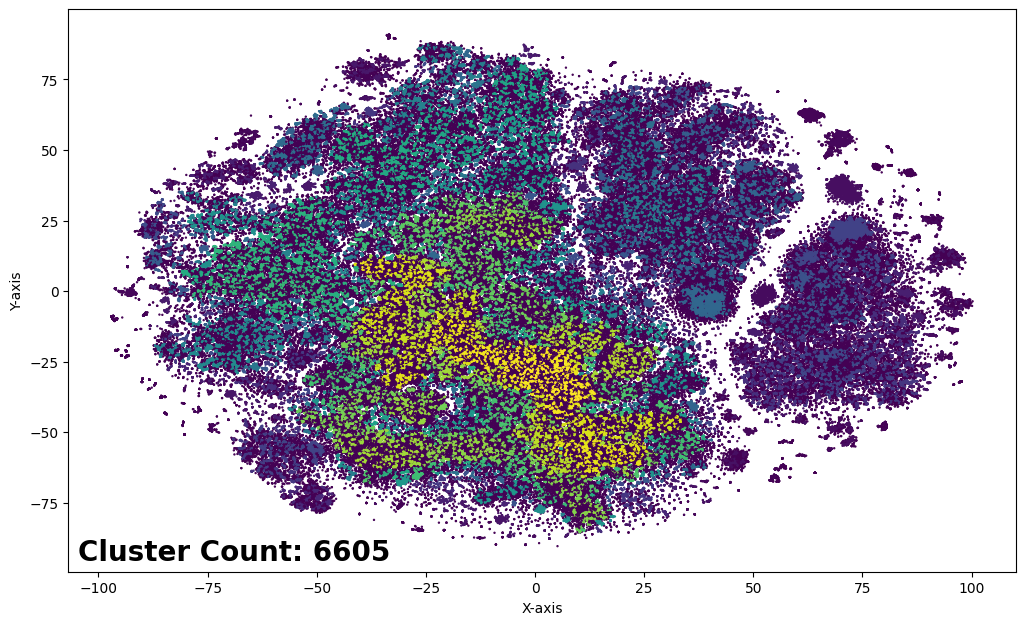}
        \caption{LARGE - Genre}
        \label{fig:large_genre}
    \end{subfigure}

    \caption{Visualizations of 2D t-SNE projections of embedded clusters on market movement, location, and genre tasks for models GPT2-SMALL, GPT2-MEDIUM, and GPT2-LARGE. Each point is a reduced and clustered headline embedding from $NIFTY$ with tags outlines in Table \ref{table:task_summary}. Each colour represent a cluster of at least 10 points. The background purple hue are points that belong to the "outlier" cluster. Results in Table~\ref{table:embedding_results} suggest larger models produce more granularity of semantic clustering.}
    \label{fig:3x3grid_apdx}
\end{figure}
\newpage
\section{Additional Details}\label{app:add-details}

\subsection{FLARE Benchmark Datasets}\label{app:flare_datasets}

\begin{table}[ht]
\centering
\caption{The dataset details in the FLARE Benchmark, reproduced here from~\cite{finma-flare-fit_xie2023pixiu} as reference.}
\resizebox{\textwidth}{!}{
\begin{tabular}{@{}llrrlll@{}}
\toprule
    Data     & Task                     & Raw   & Instruction & Data Types               & Modalities    & License     \\ \midrule
    FPB      & sentiment analysis       & 4,845 & 48,450      & news                     & text          & CC BY-SA 3.0 \\
    FiQA-SA  & sentiment analysis       & 1,173 & 11,730      & news headlines, tweets   & text          & Public       \\
    Headline & news headline classification & 11,412 & 11,412      & news headlines           & text          & CC BY-SA 3.0 \\
    NER      & named entity recognition & 1,366 & 13,660      & financial agreements     & text          & CC BY-SA 3.0 \\
    FinQA    & question answering       & 8,281 & 8,281       & earnings reports         & text, table   & MIT License  \\
    ConvFinQA & question answering       & 3,892 & 3,892       & earnings reports         & text, table   & MIT License  \\
    BigData22 & stock movement prediction & 7,164 & 7,164       & tweets, historical prices & text, time series & Public       \\
    ACL18    & stock movement prediction & 27,053 & 27,053      & tweets, historical prices & text, time series & MIT License  \\
    CIKM18   & stock movement prediction & 4,967 & 4,967       & tweets, historical prices & text, time series & Public       \\ \bottomrule
\end{tabular}
}
\end{table}

\begin{table}[H]
\caption{Example prompts for the tasks in the FLARE Benchmark, reproduced here from \cite{finma-flare-fit_xie2023pixiu} as reference}
\centering
\resizebox{\textwidth}{!}{
\begin{tabular}{ll}
\hline
Data & Prompt\\
\hline
FPB & "Analyze the sentiment of this statement extracted from a financial news article. \\
    & Provide your answer as either negative, positive or neutral. \\
    & For instance, 'The company’s stocks plummeted following the scandal.' would be classified as negative." \\
\hline
FiQA-SA & "What is the sentiment of the following financial \{category\}: \\
         & Positive, Negative, or Neutral?" \\
\hline
Headline & "Consider whether the headline mentions the price of gold. \\
         & Is there a Price or Not in the gold commodity market indicated in the news headline? \\
         & Please answer Yes or No." \\
\hline
NER & "In the sentences extracted from financial agreements in U.S. SEC filings, \\
    & identify the named entities that represent a person (`PER'), an organization (`ORG'), \\
    & or a location (`LOC'). The required answer format is: `entity name, entity type'. \\
    & For instance, in 'Elon Musk, CEO of SpaceX, announced the launch from Cape Canaveral.', \\
    & the entities would be: 'Elon Musk, PER; SpaceX, ORG; Cape Canaveral, LOC'" \\
\hline
FinQA & "Given the financial data and expert analysis, please answer this question:" \\
\hline
ConvFinQA & "In the context of this series of interconnected finance-related queries and the additional information \\
           & provided by the pretext, table data, and post text from a company’s financial filings, \\
           & please provide a response to the final question. This may require extracting information \\
           & from the context and performing mathematical calculations. Please take into account the information provided \\
           & in the preceding questions and their answers when formulating your response:" \\
\hline
BigData22 & "Analyze the information and social media posts to determine if the closing price of \{tid\} \\
           & will ascend or descend at \{point\}. Please respond with either Rise or Fall." \\
\hline
\end{tabular}
}
\label{table:your_label}
\end{table}

\paragraph{FPB(Financial Phrase Bank)} Introduced by \cite{dataset-finphrasebank-malo2014good}. It contains 14,780 example sentences from finance related news which are labelled positive, negative or neutral by experts in the field.

\paragraph{FiQA-SA} Introduced by~\citep{fin-dataset_fiqa_maia201818}. It contains a total of 1,174 examples from news headlines and tweets. Each example contains the sentence and the sentence snippet associated with the target entity, aspect, and sentiment score.
An Aspect label (Level 1)  takes on one of \textit{four} possible labels (Corporate, Economy, Market or Stock), and Level 2 Aspect label takes on one of \textit{twenty-seven} possible labels (Appointment, Risks, Dividend Policy, Financial, Legal, Volatility, Coverage, Price Action, etc.). 


\paragraph{News Headline Classification} Introduced by ~\citep{sinha2021impact}. Applicable only the the commodities market, specifically gold. 

\paragraph{NER (Named Entity Recognition)} This task aims to detect and isolate crucial financial entities such as persons, organizations and locations. In the FLARE benchmark, the authors used the FIN dataset \cite{alvarado2015domain} which includes sentences from the public financial agreements through U.S. Security and Exchange Commission(SEC) filings and manually annotated entity types from LOCATION(LOC), ORGANIZATION(ORG) and PERSON(PER). (Adopted from \citealp{finma-flare-fit_xie2023pixiu})

\paragraph{FinQA} For this task, the authors use two datasets; FinQA \cite{chen2021finqa} and ConvFinQA \cite{chen2022convfinqa}. FinQA consists of Q\&A pairs annotated by experts and their corresponding earnings reports from S\&P 500 companies. ConvFinQA is a multi-turn Q\&A version of the FinQA.

\paragraph{Stock Movement Prediction Datasets and Tasks: Flare-SM tasks} 
\textbf{FLARE} proposed by \citet{finma-flare-fit_xie2023pixiu}, extends to include one financial prediction task -- the \textbf{CIKM} dataset~\cite{fin-dataset_cikm_wu2018hybrid} as an evaluation task among (four) other general financial NLP tasks. Under the hood, this benchmark is a fork of the `\textit{lm-eval}` harness~\cite{eval-harness} with addendums. Other stock price movement prediction from social dataset include \textbf{StockNet}~\cite{fin-dataset_acl18_stocknet_xu2018stock} which is mainly stock tweets of 88 stock tickers from 9 financial market industries from Twitter over two years (from 2014-2015) aligned with their corresponding historical price data. \textbf{BigData22}~\citep{fin-dataset_bigdata22_soun2022accurate} is another more recent tweets dataset comprising of tweets about 50 stock tickers during the period 2019-07-05 to 2020-06-30.

\section{Additional Related Work}\label{app:related-work}
In this section we enclose works encompassing ML/AI/RL based techniques for financial market downstream tasks, specifically tasks pertaining to market forecasting (that can be movement prediction, or, regression tasks of price forecasting).

\subsection{History of using PLMs, then LLMs in the Financial domain}\label{app:ss:plm-history}
Many PLMs for the financial domain have been proposed by continual pre-training PLMs with large-scale financial texts. \cite{araci2019finbert} proposed the first financial PLM called FinBERT that pre-trained BERT~\citep{kenton2019bert} with open released financial corpus such as TRC2financial~\citep{nist_reuters_dataset} and Financial Phrase Bank~\citep{malo2014good}. FinBERT outperforms neural network methods such as LSTM in financial sentiment classification tasks.
\cite{yang2020finbert} further proposed FinBERT by pre-training BERT with a 4.9 billion tokens financial communication corpus, which outperforms BERT on three financial sentiment classification datasets.
\cite{shah2022flue} proposed FLANG, a financial PLM with BERT and ELECTRA~\citep{clark2020electra} as the backbone. Besides English, financial PLMs in other languages, such as Chinese, were also proposed, such as Mengzi-fin~\citep{zhang2021mengzi} and BBT-FinT5~\citep{lu2023bbt}. 

\paragraph{Financial LLM Evolution}
Latest, \cite{wu2023bloomberggpt} proposed BloombergGPT, the first financial large language model with 50 billion parameters, that is pre-trained with mixed datasets from the general and financial domain. However, neither the model nor pre-trained domain datasets are released. The model is also not instruction-following like other LLMs such as ChatGPT and GPT-4.
Meta AI's LLaMA~\citep{touvron2023llama} was the first open-source LLM with parameters ranging from 7B and 13B to 65B that gained widespread traction in the research and open-source community. LLaMA-13B has comparable and even better performance than GPT-3~\citep{brown2020language} with 175B parameters on common sense reasoning tasks. Following efforts have been proposed to improve LLaMA for instruction following like ChatGPT, by instruction tuning.
Such as  the Alpaca~\cite{alpaca} model by fine-tuning LLaMA-7B with 52K instruction-following samples generated with the self-instruct method~\citep{wang2022self}. 
\cite{vicuna2023} proposed Vicuna-13B by fine-tuning LLaMA-13B with 70K conversation data from ShareGPT~\citep{sharegpt_website}. It can generate better answers to user's questions compared with Alpaca.
However, there are no open-sourced LLMs and instruction-tuning data entirely focused on the financial domain. FinMA~\cite{finma-flare-fit_xie2023pixiu} series of model along with the recently release Flare benchmark aims to fill this void, however, these models uses (Llama 1~\cite{llama-touvron2023llama}) as the base model that were not tuned to be instruction following assistants.

\subsection{More Related Works}
We enclose further works from related literature with high-level breakdown of their key contributions in each that may be of interest to target audience at the intersection of finance, RL and downstream financial tasks.

\begin{enumerate}
    \item Financial Trading as a Game: A Deep Reinforcement Learning Approach \cite{jiang2017financial}
    \begin{itemize}
        \item Stock trading AI
        \item Stock market as a dynamic environment that can be modeled as a game
        \item Deep Q Network learns to trade from market data features
    \end{itemize}
    \item A Deep Reinforcement Learning Framework for the Financial Portfolio Management Problem \cite{jiang2017deep}
    \begin{itemize}
        \item DRL to dynamically allocate funds among a set of assets
        \item MDPs to model the portfolio management task and employs a policy gradient method to optimize the investment strategy
    \end{itemize}
    \item True Knowledge Comes from Practice: Aligning LLMs with Embodied Environments via Reinforcement Learning \cite{tan2024true}
    \begin{itemize}
        \item Uses powers of LLM knowledge base, and RL's environment alignment to make better decisions
        \item novel parameter-efficient training architecture where the actor and critic share one frozen LLM equipped with low-rank adapters (LoRA) updated by PPO
    \end{itemize}
    \item Stock Market Prediction Using Deep Reinforcement Learning \cite{stockmarket-pred-awad-2023}
    \begin{itemize}
        \item Introduction of a New Framework: Proposes a combined architecture leveraging ANN, LSTM, NLP, and DRL techniques for predicting stock market trends, specifically focusing on gold stocks.
        \item Utilization of Sentiment Analysis: Employs natural language processing to process news and social media data, enhancing the prediction accuracy through sentiment analysis.
        \item Incorporation of Historical Data: Uses historical stock price data from major platforms like SandP, Yahoo, and NASDAQ to inform the predictive model.
        \item Application of LSTM and VMD: Applies LSTM networks for price prediction and Variational Mode Decomposition (VMD) for signal processing, improving prediction reliability.

        \item Innovative Use of BERT and TF-IDF: Enhances sentiment analysis phase by fine-tuning BERT models with TF-IDF for maximum accuracy in interpreting financial news sentiment.
        \item Conclusive Evidence of Efficacy: Provides conclusive results showing the effectiveness of the integrated approach in predicting stock market trends, particularly for gold stocks, with high accuracy and improved profitability potential.
    \end{itemize}
    \item LLM-Informed Multi-Armed Bandit Strategies for Non-Stationary Environments \cite{decurto2023llminformed}
    \begin{itemize}
        \item innovative strategy for the multi-armed bandit problem in dynamic environments by integrating large language models (LLMs) to guide decision-making.
        \item nparameter-efficient architecture combining LLMs with reinforcement learning to optimize the balance between exploration and exploitation.
    \end{itemize}
    \item Temporal Data Meets LLM -- Explainable Financial Time Series Forecasting(\cite{xinli-yu-temporal-data-meets-llm})
        \begin{itemize}
        \item Introduction to LLM in Finance: Investigates LLMs' capability for explainable financial forecasting, addressing challenges like cross-sequence reasoning and multi-modal signal integration.
        \item Methodology: Utilizes NASDAQ-100 stock data, company metadata, and economic/financial news for LLM-based forecasting, employing GPT-4 and Open LLaMA models.
        \item Experiments with GPT-4 and Open LLaMA: Demonstrates zero-shot/few-shot inference and fine-tuning techniques to enhance forecasting performance.
        \item Superior Performance Over Traditional Models: Shows that LLM approaches, particularly GPT-4 with Chain of Thought (COT), outperform traditional ARMA-GARCH and gradient-boosting tree models in accuracy and explanation quality.
        \item Future Directions: Suggests further research into extending studies to other stock indexes, integrating more data types, and exploring fine-tuning of larger LLMs for enhanced reasoning capabilities.
    \end{itemize}
    \item Unveiling the Potential of Sentiment: Can Large Language Models Predict Chinese Stock Price Movements?\cite{zhang2023unveiling}
    \begin{itemize}
        \item Benchmark and Framework Development: The authors introduce a comprehensive benchmark and a standardized back-testing framework to objectively assess the performance of various LLMs in extracting sentiment factors from Chinese financial news texts.
        \item Model Comparison: Three types of models are compared: generative LLM (ChatGPT), Chinese language-specific pre-trained LLM (Erlangshen-RoBERTa), and financial domain-specific fine-tuned LLM classifier (Chinese FinBERT).
        \item Sentiment Extraction and Trading Strategy: The study involves extracting sentiment factors from a large volume of Chinese news summaries and constructing quantitative trading strategies to evaluate the models' performance through back-tests.
        \item Results: The Erlangshen-RoBERTa model outperforms the others in terms of annual return, risk-adjusted return, and excess return, demonstrating the importance of language-specific pre-training and fine-tuning in sentiment analysis for the Chinese stock market.
        \item Conclusions: The research highlights the potential of LLMs in enhancing quantitative stock trading strategies by leveraging sentiment analysis, emphasizing the effectiveness of language-specific models and methodologies over general model size for Chinese financial texts.
    \end{itemize}
    \item Reinforcement Learning for Optimizing RAG for Domain Chatbots \cite{kulkarni2024reinforcement}
    \begin{itemize}
        \item The paper presents a method to optimize Retrieval Augmented Generation (RAG) for domain chatbots by using Reinforcement Learning (RL) to reduce the number of tokens required from a Large Language Model (LLM), thus saving costs while maintaining or slightly improving accuracy.
        \item It introduces a policy-based model that decides whether to fetch FAQ context for a query or not, demonstrating significant cost savings (~31percent) and improved retrieval accuracy through experimental results.
    \end{itemize}
\item TradingGPT: Multi-Agent System with Layered Memory and Distinct Characters for Enhanced Financial Trading Performance
\cite{li2023tradinggpt}
\begin{itemize}
    \item Introduces a multi-agent framework utilizing Large Language Models (LLMs) with layered memories to improve financial trading decisions, aligning closer to human memory processes.
    \item Proposes a novel method where trading agents are equipped with individualized characters and risk preferences to diversify trading strategies and enhance market opportunity identification.
    \item Incorporates real-time multi-modal data processing for comprehensive financial analysis, enabling agents to adapt quickly to market changes for both daily and high-frequency trading.
    \item Details the system architecture, including memory formulation based on individual and inter-agent experiences, and the design of training and testing workflows to optimize trading strategies.
    \item Demonstrates potential for superior trading performance through the simulation of realistic trading scenarios, aiming for future application in various domains beyond finance, like gaming and healthcare.
\end{itemize}
    
\end{enumerate}





\end{document}